\documentclass[showpacs,floats,eqsecnum,prb]{revtex4}
\usepackage{amssymb}

\usepackage{epsfig}
\usepackage{epstopdf}
\usepackage{graphicx}

\begin{document}

\title{Electronic transport through ferromagnetic and superconducting
junctions with spin-filter tunneling barriers.}
\author{F. S. Bergeret$^{1,2,3}$, A. Verso$^{2}$ and A. F. Volkov$^{2,4}$ }
\affiliation{$^1$ Centro de F\'{\i}sica de Materiales (CFM-MPC), Centro Mixto
CSIC-UPV/EHU, Manuel de Lardizabal 4, E-20018 San Sebasti\'{a}n, Spain\\
$^2$Donostia International Physics Center (DIPC), Manuel de Lardizabal 5,
E-20018 San Sebasti\'{a}n, Spain\\
$^3$ Institut f\"ur Physik, Carl von Ossietzky Universit\"at, D-26111
Oldenburg, Germany\\
$^4$Theoretische Physik III, Ruhr-Universit\"{a}t Bochum, D-44780 Bochum,
Germany\\
}
\date{\today }

\begin{abstract}
$\bar{t}$We present a theoretical study of the quasiparticle and subgap
conductance of generic $X/I_{sf}/S_{M}$ junction with a spin-filter barrier $%
I_{sf}$, where $X$ is either a normal $N$ or a ferromagnetic metal $F$ and $%
S_{M}$ is a superconductor with a built-in exchange field. Our study is
based on the tunneling Hamiltonian and the Green's function technique.
First, we focus on the quasiparticle transport, both above and below the
superconducting critical temperature. We obtain a general expression for the
tunneling conductance which is valid for arbitrary values of the exchange
field and arbitrary magnetization directions in the electrodes and in the
spin-filter barrier. In {the }second part we consider the subgap conductance
of a normal metal-superconductor junction with a spin-filter barrier. We
provide a heuristic derivation of new boundary conditions for the
quasiclassical Green's functions which take into account the spin-filter
effect at interfaces. With the help of these boundary conditions, we show
how the proximity effect and the subgap conductance are suppressed by
spin-filtering in a $N/I_{sf}/S$ junction, where $N$ is a normal metal. Our
work provides useful tools for the study of spin-polarized transport in
hybrid structures both in the normal and in the superconducting state.
\end{abstract}

\maketitle

%\affiliation{Theoretische Physik III,\\
%Ruhr-Universit\"{a}t Bochum, D-44780 Bochum, Germany\\
%}

\bigskip

\section{Introduction}

Over the last decade, there has been a growing interest in studying
superconductor/ferromagnet ($S/F$) hybrid structures. On the one hand, this
interest is due to the progress in technology that allows a controllable
fabrication of nanohybrid structures using a wide range of superconducting
and magnetic materials. On the other hand, this interest is due to the
discovery of new and interesting fundamental phenomena, as for example the
so-called $\pi -$state in $S/F/S$ junctions Josephson\cite%
{Bulaev77,Buzdin82,Ryaz,Kontos,Blum,Bauer,Sellier,PalVolkovEfetov,Weides09},
and more recently the long-range proximity effect mediated by {odd-frequency
triplet} superconducting correlations in S/F structures \cite%
{BVE01,Kaizer06,Sosnin,Birge,Westerholt10,Chan,BlamireScience,Aarts} (for an
overview see Refs.\cite{GolubovRMP,BuzdinRMP,BVErmp,EschrigPhysToday,Zabel}).

The triplet superconducting correlations can carry spin-polarized
supercurrents, \textit{i.e.} currents without dissipation, that can be
exploited in several ways in spintronics devices\cite{EschrigPhysToday}. In
this context, the use of tunnel barriers with spin-dependent transmission,
the so-called spin-filters, may be desirable for the creation of such spin
supercurrents. Spin-filters are tunnel barriers with spin-dependent barrier
height. They have been used for decades to generate polarized currents in
spintronic circuits \cite{Moodera90,Moodera08}.

In spite of numerous works devoted to the theoretical study of $S/F$
structures, the study of the spin-filter effect in connection {with} the
transport properties of S/F structures still remains open. For example, in
Ref.\cite{Valles} the transport properties of an S/F junction was calculated
by using the Blonder-Tinkham-Klapwijk formalism \cite{Blonder}. This
analysis was extend in other several works\cite%
{Linder,Linder2,Fogelstrom2000, Kalenkov09,Tanaka97,Kawabata10, Belzig09}
for S/F and S/F/S junctions in the ballistic and diffusive limit by taking
into account spin-active interfaces between the F and S layers\cite%
{Zaitsev84,Millis88}. In particular, the results for the diffusive limit
presented in those works have been obtained by using the boundary conditions
for the quasiclassical Green functions derived in \cite{Belzig09}. However,
as we will show in section IV, these boundary conditions cannot describe the
spin-filter effects and hence, none of the above mentioned works addressed
the question how the spin-filtering affects the proximity effect in S/F
structures. Only recently, we have analyzed\cite{Bergeret12} the effect of
spin filtering on the Josephson current through a $SF/I_{sf}/FS$ junction.
It was shown that even in the case of a highly spin-polarizing barrier, a
Josephson junction can flow provided the magnetizations of the F layers are
non collinear. The results of Ref.\cite{Bergeret12} have been obtained from
a model that combines the tunneling Hamiltonian and the quasiclassical
Green's functions, and provide a plausible explanation for a recent
experiment on spin-filter Josephson junctions\cite{Blamire}. Note however,
that the model used in Ref.\cite{Bergeret12} assumes exchange fields to be
smaller than the Fermi energy and therefore cannot be straightforwardly
generalized for the case of strong ferromagnets.

In the current paper, we present a general theory for the conductance
through different hybrid structures with spin-filters as barriers, arbitrary
values of the exchange field and arbitrary directions of the magnetization
in the barrier and in the electrodes. We start with the model used in Ref.%
\cite{Bergeret12} and extend it to dissipative tunnel junctions. In the
first part we focus on the study of the quasi-particle current and derive a
general expression for the tunneling conductance. This expression recovers
well known results in particular limiting cases and predicts new effects
related to the mutual orientation of the magnetizations. We study the
tunneling conductance of different junctions like $F/I_{sf}/F$, $%
HM/I_{sf}/HM $, and $F/I_{sf}/S$ ($HM$ stands for a ferromagnetic
half-metal). In the second part we focus on the subgap transport through a $%
N/I_{sf}/S$ junction using the quasiclassical formalism. In order to
quantify the effect of spin-filtering on the proximity effect we need to
generalize the existing boundary conditions\cite{K-L,Belzig09} (BCs) for the
quasiclassical equations. Accordingly, we present a heuristic derivation of
new BCs which account for the spin-filter effect. These boundary conditions
can be used in a wide range of problems involving superconductors,
ferromagnets and spin-filter tunnel barriers. As an example, we study the
subgap conductance of an $N/I_{sf}/S$ junction and show its suppression due
to the spin-filter effect. Thus, our work provides on the one hand a new
powerful tool for the theoretical study of spin transport in hybrid
structures and on the other hand general expressions for the conductance
that can be used for the interpretation of a broad range of experiments on
spin transport through spin-filters.

\section{Model}

\begin{figure}[tb]
\includegraphics[width=0.5\columnwidth]{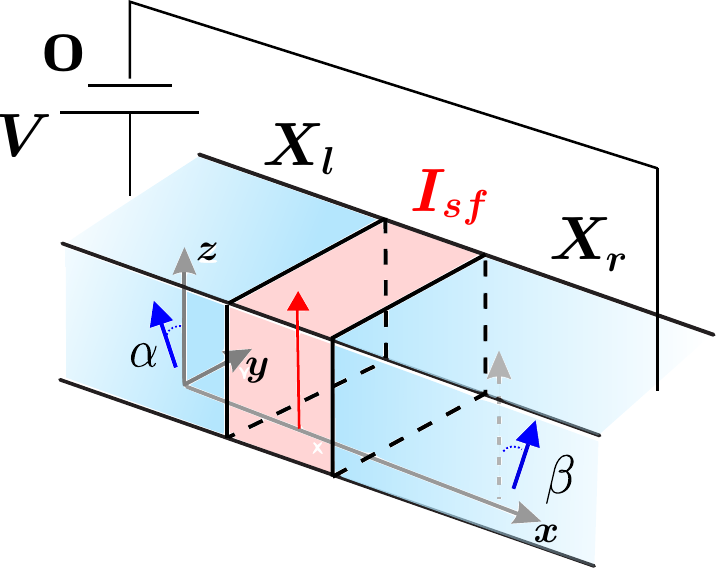} \vspace{-2mm}
\caption{A generic tunnel junction $X_{l}/I_{sf}/X_{r}$. The left and right
electrodes, $X_{l}$ and $X_{r}$, are either a ferromagnet $F$, normal metal $%
N$ or a superconductor $S_{M}$ with a built-in exchange field $h$. The layer
$I_{sf}$ is a spin-filter barrier. $\protect\alpha $ and $\protect\beta $
are the orientation of the exchange field in the left and right electrodes.}
\label{fig0}
\end{figure}

We consider two type of junctions, first a generic tunnel junction $%
X_{l}/I_{sf}/X_{r}$ as the one shown in Fig. \ref{fig0}. The left and right
electrodes, $X_{l}$ and $X_{r}$ are either ferromagnetic $F$ or a
superconductor $S_{M}$ with a built-in exchange field $h$. {The layer }$%
I_{sf}$ is a spin-filter barrier, \textit{i.e.,} a spin-dependent tunneling
barrier. The second type of junction we considered, is a $N/I_{sf}/S$
spin-filter junction between a normal metal ($N$) and a conventional
superconductor ($S$). Our first aim is to derive a general expression for
the current in both structures. For this purpose, we use the well known
tunneling Hamiltonian method, which has been used in several works on
tunneling in {superconducting junctions\ (see the books \cite%
{KulikBook,BaroneBook,Schrieffer,Wolf85} and references therein) and in
systems with charge- and spin-density waves} \cite{Artemenko83,Moor12}. The
Hamiltonian consists of the Hamiltonians of the left (right) electrodes and
the tunneling term %%%%%%%%%%%%%%%
%%%%%%%%%%%%%%%%%%FIGURE 1%%%%%%%%%%%%%%%%%%
%%%%%%%%%%%%%%%
%%%%%%%%%%%%%%%%%%%%
\begin{equation}
H=H_{r}+H_{l}+H_{T}\;.  \label{Hamiltonian}
\end{equation}%
For the electrodes we consider the general Hamiltonian
\begin{equation}
H_{l}=\sum_{\mathbf{p},s}\xi _{\mathbf{p}}a_{\mathbf{p}s}^{\dagger }a_{%
\mathbf{p}s}+\sum_{\mathbf{p}}\left( \Delta a_{\mathbf{p}\uparrow }^{\dagger
}a_{-\mathbf{p}\downarrow }^{\dagger }+h.c.\right) -\sum_{\mathbf{p}%
,s,s^{\prime }}a_{\mathbf{p}s}^{\dagger }(h_{l}\mathbf{n.\hat{\sigma}}%
)_{ss^{\prime }}a_{\mathbf{p}s^{\prime }},  \label{HamRL}
\end{equation}%
where $a_{\mathbf{p}s}(a_{\mathbf{p}s}^{\dagger })$ is the annihilation
(creation) operator of a particle with momentum $\mathbf{p}$ and spin $s$, $%
\xi _{\mathbf{p}}$ is the quasiparticle energy, $\Delta $ the
superconducting gap, $\mathbf{\sigma }=(\sigma _{1},\sigma _{2},\sigma _{3})$
is a vector with the Pauli matrices, $h_{l}$ the amplitude of the effective
exchange field and $\mathbf{n}$ a unit vector pointing in its direction.
Similar Hamiltonian can be written for the right electrode.

The $H_{T}$ term in Eq. (\ref{Hamiltonian}) describes the spin-selective
tunneling through the spin-filter barrier $I_{sf}$ and is given by
\begin{equation}
H_{T}=\sum_{\{n,s,\mathbf{p,p}^{\prime }\}}\left( \mathcal{T}\hat{\sigma}%
_{0}+\mathcal{U}\hat{\sigma}_{3}\right) _{ss^{\prime }}a_{\mathbf{p}%
s}^{\dagger }b_{\mathbf{p}^{\prime }s^{\prime }}+h.c.  \label{HamTun}
\end{equation}%
where $a$ and $b$ are the operators in the left and right electrodes
respectively, $\mathcal{T}$ and $\mathcal{U}$ are the tunneling spin
independent and spin dependent matrix elements, respectively. For simplicity
we neglect the momentum dependence of $\mathcal{T}$ and $\mathcal{U}$,
assuming that the characteristic energies are much smaller than the Fermi
energy and that by tunneling the electrons are scattered randomly. The
tunneling amplitude for spin up(down) is given by the relation: $T_{\uparrow
(\downarrow )}=\mathcal{T}\pm \mathcal{U}$. The origin of the different
tunneling amplitudes might be, for example, the conduction-band splitting in
the ferromagnetic semiconducting barrier which leads to different tunnel
barrier heights for spin-up and spin-down electrons \cite%
{Moodera90,Moodera08}.

In order to write the equation of motions for the Green's functions it is
convenient to write the tunneling Hamiltonian (\ref{HamTun}) in terms of new
operators defined in an enlarged space (spin$\otimes $particle-hole). These
operators are defined as:
\begin{equation}
A_{n,s}=\left\{
\begin{array}{c}
a_{s}\;\;{\text{f}or}\;\;n=1 \\
a_{\bar{s}}^{\dagger }\;\;{\text{f}or}\;\;n=2%
\end{array}%
\right.
\end{equation}%
where $s=1,2$ is the spin up (down) index and $\bar{s}$ implies the change $%
1\leftrightarrow 2$. In analogy one introduces the operators $B_{n,s}$ for
the right electrode. By using these operators the tunneling term (\ref%
{HamTun}) can be written as
\begin{equation}
H_{T}=\sum_{\{n,s,\mathbf{p,p}^{\prime }\}}[\check{A}_{\mathbf{p}}^{\dagger
}\left( \mathcal{T}\hat{\tau}_{3}\otimes \hat{\sigma}_{0}+\mathcal{U}\hat{%
\tau}_{0}\otimes \hat{\sigma}_{3}\right) \check{B}_{\mathbf{p}^{\prime
}}+h.c.]  \label{HamTun1}
\end{equation}%
where $\check{A}=A_{n,s}$ and $\check{B}=B_{n,s}$. The Hamiltonian (\ref%
{HamRL}) transforms to %$\check{A}$ and $\check{B}$
(see, for example, Ref.\cite{BVErmp})
\begin{equation}
H_{l}=\sum_{\{n,s,\mathbf{p}\}}\check{A}_{\mathbf{p}}^{\dagger }\mathcal{H}%
_{l}\check{A}_{\mathbf{p}},  \label{HamRL1}
\end{equation}%
with $\mathcal{H}_{l}=\xi _{\mathbf{p}}\hat{\tau}_{3}\otimes \hat{\sigma}%
_{0}+\left( \Delta \hat{\tau}_{1}\otimes \hat{\sigma}_{3}+h.c.\right) -h_{l}%
\hat{\tau}_{0}\otimes \hat{\sigma}_{3}$, for a exchange field directed along
the $z$-axis. One can take into account an arbitrary direction of the
exchange field (or magnetization vector\footnote{%
We assume as usual that the exchange field vector is parallel to the
magnetization one.}) by means of a rotation in spin-space. We assume
throughout this work that the transport takes place in the $x$-direction
while the magnetization vector of the ferromagnets lies in the junction
plane, i. e., the $(y,z)$-plane. A rotation in plane is described by the
matrix
\begin{equation}
\check{R}_{\alpha }=\cos (\alpha /2)+i\hat{\tau}_{3}\otimes \hat{\sigma}%
_{1}\sin (\alpha /2)\;,  \label{rotation}
\end{equation}%
where $\alpha $ is the rotation angle.

In order to calculate the tunneling current through the generic junction, we
first write the Dyson equation for the Keldysh Green's functions $\check{G}%
_{l}$, for instance, in the left electrode
\begin{equation}
i\partial _{t}\check{G}_{l}=1+\check{\Sigma}_{l}\check{G}_{l}+\check{%
\mathcal{H}}_{l}\check{G}_{l}  \label{Dyson}
\end{equation}%
Here $\check{\Sigma}_{l}=\sum_{\mathbf{q}}\check{\Gamma}\check{G}_{r}(%
\mathbf{q})\check{\Gamma}^{\dagger }$ is the self-energy part\cite%
{Schrieffer,Volkov75,Artemenko83,Bergeret2005,Moor12} related to the tunnel
Hamiltonian (\ref{HamTun1}), $\check{G}_{r}$ is the full
(non-quasiclassical) Green's function in the right electrode and $\check{%
\Gamma}=\mathcal{T}\hat{\tau}_{3}\otimes \hat{\sigma}_{0}+\mathcal{U}\hat{%
\tau}_{0}\otimes \hat{\sigma}_{3}.$ In the following we restrict our
analysis to the lowest order in tunneling. In this case the Green's function
$\check{G}_{l}$ is determined by Eq.(\ref{Dyson}) after neglecting the
second term on the r.h.s. The exact form of $\check{G}_{l}$ is given in Eq.(%
\ref{Gfunction}).

We proceed as usual subtracting from the Eq.(\ref{Dyson}) (prelimenary
multiplied by $\hat{\tau}_{3}$\ from the left) its conjugated equation
multiplied by $\hat{\tau}_{3}$ from from the right:
\begin{equation}
i(\hat{\tau}_{3}\partial _{t}\check{G}+\partial _{t^{\prime }}\check{G}\hat{%
\tau}_{3})_{l}=\hat{\tau}_{3}\check{\Sigma}\check{G}-\check{\check{G}\check{%
\Sigma}\hat{\tau}_{3}+\hat{\tau}_{3}\check{G}\check{\mathcal{H}}_{l}-%
\mathcal{H}}_{l}\check{G}\hat{\tau}_{3}|_{l}\;.  \label{Dyson1}
\end{equation}%
If we now multiply the Keldysh component of this equation by the electron
charge $e,$ set $t=t^{\prime }$, take the trace and sum up over momenta, we
obtain on the l.h.s the time derivative of the charge, $\partial _{t}Q$.
Thus, the current density $j$ through the barrier is then determined by the
first two terms in the r.h.s.
\begin{equation}
I_{T}=-\frac{e}{32\pi }\sum_{\mathbf{p,q}}\int \frac{d\epsilon }{2\pi }%
\mathrm{Tr}\left\{ \hat{\tau}_{3}\otimes \hat{\sigma}_{0}\left[ \check{\Gamma%
}_{\alpha \beta }\check{G}_{r}(\mathbf{q},\epsilon )\check{\Gamma}_{\alpha
\beta }^{\dagger },\check{G}_{l}(\mathbf{p},\epsilon )\right] ^{K}\right\}
\label{Current1}
\end{equation}%
where $\check{\Gamma}_{\alpha \beta }\equiv \check{R}_{\alpha }^{\dagger }\check{\Gamma}%
\check{R}_{\beta }$, and we have defined $\check{\Gamma}$ %
after Eq.(\ref{Dyson}). The Green's functions $\check{G}_{l,r}$ correspond
to the case of the magnetization vector oriented parallel to the $z$-axis.
We expressed the Green's functions $\check{G}_{l\alpha }(\mathbf{p},\epsilon
)$ for arbitrary magnetization orientation through the matrices $\check{G}%
_{l,r}$ as follows: $\check{G}_{l\alpha }(\mathbf{p},\epsilon )=\check{R}%
_{\alpha }\int d(t-t^{\prime })$ $\check{G}_{l0}(\mathbf{p},t-t^{\prime
})\exp (i\epsilon (t-t^{\prime }))\check{R}_{\alpha }^{\dagger }.$

%One can represent the matrix $\check{\Gamma}_{\alpha \beta },$ which
%describes the tunneling, in the form $\check{\Gamma}_{\alpha \beta }\equiv
%\check{A}_{\alpha \beta }+\check{B}_{\alpha \beta }$ with $\check{A}_{\alpha
%\beta }=\mathcal{T}\cos ((\alpha -\beta )/2)\hat{\tau}_{0}\otimes \hat{\sigma%
%}_{0}+\mathcal{U}\cos ((\alpha +\beta )/2)\hat{\tau}_{3}\otimes \hat{\sigma}%
%_{3}$ and $\check{B}_{\alpha \beta }=i\mathcal{T}\sin ((\alpha -\beta )/2)%
%\hat{\tau}_{3}\otimes \hat{\sigma}_{1}-\mathcal{U}\sin ((\alpha +\beta )/2)%
%\hat{\tau}_{0}\otimes \hat{\sigma}_{2}.$ Note that the matrix $\check{A}%
%_{\alpha \beta }$ commutes with the matrices $\hat{\sigma}_{0}$ and $\hat{%
%\sigma}_{3},$ whereas the matrix $\check{B}_{\alpha \beta }$ anticommuts
%with the matrix $\hat{\sigma}_{3}.$ This property is useful for the
%calculation of the current density.

In the case that the energies involved in the problem are much smaller than
the Fermi energy, one can perform the momentum integration in Eq.(\ref%
{Current1}) and the current can be written in terms of quasiclassical\
Green's functions $\check{g}_{l\alpha }=(i/\pi )\hat{\tau}_{3}\otimes \hat{%
\sigma}_{0}\int d\xi _{p}\check{G}_{l\alpha }(\mathbf{p},\epsilon )$
\begin{equation}
I_{T}R_{N}=[16e(\mathcal{T}^{2}+\mathcal{U}^{2})]^{-1}\int d\epsilon \mathrm{%
Tr}\left\{ \hat{\tau}_{3}\otimes \hat{\sigma}_{0}\left[ \mathit{\check{\Gamma%
}}_{\alpha \beta }\,\check{g}_{r}(\epsilon )\mathit{\check{\Gamma}}_{\alpha
\beta }^{\dagger },\check{g}_{l}(\epsilon )\right] ^{K}\right\}
\label{Current2}
\end{equation}%
where $\mathit{\check{\Gamma}}_{\alpha \beta }=\check{\Gamma}_{\alpha \beta }\hat{\tau}_{3}\otimes
\hat{\sigma}_{0}%=\mathcal{T}+\mathcal{U}\hat{\tau}_{3}\otimes \hat{\sigma}_{3}
.$ The resistance $R_{N}=[4\pi e^{2}N_{l}(0)N_{r}(0)(\mathcal{T}^{2}+%
\mathcal{U}^{2})]^{-1}$ is the junction resistance in the normal state, i.
e., the resistance of an $F/I_{sf}/F$ junction with parallel orientation of
magnetization along the $z$-axis, $N_{l,r}(0)=(p_{F}m/2\pi ^{2})_{l,r}$ are
the density of states (DOS) at the Fermi level. One should have in mind that
by going over to the quasiclassical Green's functions we lose the spin
dependence of the DOS {in the normal state}. In that case the retarded
(advanced) Green's functions $\check{g}_{l,r}^{R(A)}$ in the ferromagnet
have a trivial structure in spin-space, $\check{g}_{F}^{R(A)}=\pm \hat{\tau}%
_{3}\otimes \hat{\sigma}_{0}$, so that the normalized density of states is
the same for spin up and down. This approach is valid for electrodes with
small spin-splitting at the Fermi level, and was used for example in Ref.%
\cite{Bergeret12} for the calculation of the Josephson current through a $%
S/I_{sf}/S$ junction. However, if the spin-polarization of the electrodes at
the Fermi level is large enough one has to use Eq.(\ref{Current1}) in order
to compute the current. This is done in the next section, where we calculate
the conductance of a $F/I_{sf}/S_{M}$ and a $F/I_{sf}/F$ junction.

\section{The conductance for junctions with arbitrary exchange fields}

In this section we consider junctions of the type $F/I_{sf}/F$ and $%
F/I_{sf}/S_{M}$, where F is a ferromagnet and $S_{M}$ describes either a
thin $FS$ bilayer\cite{Bergeret2001a} or a superconductor with an induced
spin-splitting field due to the proximity of the magnetic barrier $I_{sf}$%
\cite{Sauls88a}. We are interested in arbitrary strength of exchange field
and therefore we have to go beyond quasiclassics and use Eq. (\ref{Current1}%
) for the current. We assume a bias voltage $V$ between the electrodes {%
setting the electric potential in the superconducting electrode equal to
zero (Fig. \ref{fig0})}. In the tunneling limit the junction under
consideration can only carry a normal (quasiparticle) current, which is
determined by the normal Green's functions $\check{G}_{l,r}$. These are
diagonal in spin space with diagonal elements given by
\begin{equation}
\hat{G}_{r\pm }^{R}(\mathbf{p})=\frac{(\epsilon _{\pm }+i\gamma )\hat{\tau}%
_{0}+\xi _{\mathbf{p}}\hat{\tau}_{3}}{(\epsilon _{\pm }+i\gamma )^{2}-(\xi _{%
\mathbf{p}}^{2}+\Delta ^{2})}\text{ }  \label{Gfunction}
\end{equation}%
where $\epsilon _{\pm }=\epsilon \pm h_{r},$ $\xi _{\mathbf{p}%
}=(p^{2}-p_{F,r}^{2})/2m_{r}$ and $\gamma $ is a damping in the excitation
spectrum of the superconductor due to inelastic processes or {due to
coupling with the normal metal electrode}. The corresponding Green's
function in the left ($F$) electrode has the same form if we set $\Delta =0$
and replace the index $r$ by $l$. As usual, the advanced Green's function $%
G^{A}$ is defined in a similar way with the opposite sign of the damping
term, $-i\gamma $. The full Green's function in a superconductor has the
form $\check{G}_{r}^{R}=\hat{G}_{r}^{R}\otimes \hat{\tau}_{0}+\hat{F}%
_{r}^{R}\otimes \hat{\tau}_{1},$ where $\hat{G}_{r}^{R}=G_{r0}^{R}\hat{\sigma%
}_{0}+G_{r3}^{R}\hat{\sigma}_{3},$ $G_{r0,3}^{R}=(G_{r+}^{R}\pm
G_{r-}^{R})/2,$ and $\hat{F}_{r}^{R}$ is the anomalous (Gor'kov's) Green's
function. Using the fact that the normal part of matrices $\check{G}_{r,l}$
are diagonal in the spin and particle-hole space, we can represent the
current, Eq. (\ref{Current1}) in the form
\begin{equation}
I_{T}=\frac{\pi e}{2}[(N_{n\uparrow }+N_{n\downarrow })_{l}(N_{n\uparrow
}+N_{n\downarrow })_{r}]\int d\epsilon n_{V}(\epsilon )\mathrm{Tr}\left\{
\hat{\tau}_{0}\otimes \hat{\sigma}_{0}\left[ \check{\Gamma}_{\alpha \beta }\,%
\check{\nu}_{l}(\epsilon )\check{\Gamma}_{\alpha \beta }^{\dagger }\,\check{%
\nu}_{r}(\epsilon )\right] \right\}   \label{CurrentGen}
\end{equation}%
where $n_{V}(\epsilon )=\{\tanh [(\epsilon +eV)/2T]-\tanh [(\epsilon
-eV)/2T]\}/2,$ and $T$ is the temperature. The matrices $\check{\nu}%
_{l,r}(\epsilon )$ are related to $\hat{G}_{r\pm }^{R}(\mathbf{p})$ via $%
\check{\nu}(\epsilon )=(i/2\pi ){(N_{0\uparrow }+N_{0\downarrow })}%
^{-1}\sum_{\mathbf{p}}[\check{G}^{R}(\mathbf{p})-\check{G}^{A}(\mathbf{p})]$%
, and can be written in the form: $\check{\nu}_{l,r\pm }(\epsilon )=[\nu
_{0}(\epsilon )\hat{\tau}_{0}\otimes \hat{\sigma}_{0}+\nu _{3}(\epsilon )%
\hat{\tau}_{3}\otimes \hat{\sigma}_{3}]_{l,r}$ with $\hat{\nu}%
_{0,3}(\epsilon )=[\hat{\nu}(\epsilon +h)\pm \hat{\nu}(\epsilon -h)]/2$. It
is useful to write the coefficients $\nu _{0,3}(\epsilon )$ in terms of of
the DOS for spin up and down, $N_{\uparrow ,\downarrow }$: $\nu
_{0,3}=(N_{\uparrow }\pm N_{\downarrow })/(N_{n\uparrow }+N_{n\downarrow })$%
, where $N_{n\uparrow ,\downarrow }(0)$ are the DOS at the Fermi level in
the normal state of ferromagnets. The matrices $\check{\Gamma}_{\alpha \beta
}$\ which describe the tunneling probability are given by

\begin{equation}
\check{\Gamma}_{\alpha \beta }=\mathcal{T}\hat{\tau}_{3}\otimes \hat{\sigma}%
_{0}\cos (\frac{\alpha -\beta }{2})+\mathcal{U}\hat{\tau}_{0}\otimes \hat{%
\sigma}_{3}\cos (\frac{\alpha +\beta }{2})+i\mathcal{T}\hat{\tau}_{0}\otimes
\hat{\sigma}_{1}\sin (\frac{\alpha -\beta }{2})-\mathcal{U}\hat{\tau}%
_{3}\otimes \hat{\sigma}_{2}\sin (\frac{\alpha +\beta }{2})  \label{Gamma}
\end{equation}

Substituting these expressions into Eq.(\ref{CurrentGen}), we find for the
normalized conductance
\begin{equation}
\mathit{G}_{\alpha \beta }(V)\equiv R_{F}dI_{T}/dV=(1/2e)\int d\epsilon
(dn_{V}/dV)Y_{\alpha \beta }(\epsilon )\;,  \label{Conductance}
\end{equation}%
where the spectral conductance $Y_{\alpha \beta }(\epsilon )$ is defined as

\begin{equation}
Y_{\alpha \beta }(\epsilon )=\{\nu _{0l}\nu _{0r}+\frac{[\mathcal{T}^{2}\cos
(\alpha -\beta )+\mathcal{U}^{2}\cos (\alpha +\beta )]}{(\mathcal{T}^{2}+%
\mathcal{U}^{2})}\nu _{3l}\nu _{3r}+2\frac{\mathcal{TU}}{(\mathcal{T}^{2}+%
\mathcal{U}^{2})}\mathcal{[}\nu _{0l}\nu _{3r}\cos \beta +\nu _{3l}\nu
_{0r}\cos \alpha ]\}\;.  \label{Y(V)}
\end{equation}%
and the resistance $R_{F}$\ is defined as\textbf{\ }$R_{F}=[\pi
e^{2}(N_{n\uparrow }+N_{n\downarrow })_{l}(N_{n\uparrow }+N_{n\downarrow
})_{r}]^{-1}.$

%
%\begin{equation}
%Y_{\alpha \beta }(\epsilon )=1+P_lP_r\frac{[\mathcal{T}^{2}\cos
%(\alpha -\beta )+\mathcal{U}^{2}\cos (\alpha +\beta )]}{(\mathcal{T}^{2}+%
%\mathcal{U}^{2})}+2\frac{\mathcal{TU}}{(\mathcal{T}^{2}+%
%\mathcal{U}^{2})}\mathcal{[}P_r\cos \beta +P_l\cos \alpha ]  \label{Y(V)Sp}\; ,
%\end{equation}
Equations (\ref{Conductance}-\ref{Y(V)}) are one of the main results of the
present paper. They determine the conductance of the generic junction of
Fig. \ref{fig0} in a quite general situation, since they are valid for
arbitrary exchange field, spin-filter strength and angles $\alpha $ and $%
\beta $. At low temperatures ($T\ll \Delta $) one can evaluate the energy
integral in Eq. (\ref{Conductance}) obtaining a simple expression for the
normalized conductance
\begin{equation}
\mathit{G}_{\alpha \beta }(V)=\frac{Y_{\alpha \beta }(eV)+Y_{\alpha \beta
}(-eV)}{2}\;.
\end{equation}%
We now proceed to consider different types of junctions and calculate the
conductance with the help of the last expressions.

%\section{Transport through F/I$_{sf}$/F and F/I$_{sf}$/S$_{M}$ Junctions.}

\subsection{Junctions with non-superconducting electrodes}

Let us first consider junctions in the normal state. For example in a $%
N/I_{sf}/F$ junction the left electrode is a normal metal with no
spin-polarization, therefore $\nu _{3}=0$ and $\nu _{0}=1$. From Eq.(\ref%
{Y(V)Sp}) we then obtain
\begin{equation}
Y_{\alpha \beta }^{(NF)}=1+P_{b}P_{r}\cos \beta \;,  \label{Y(V)Sp1}
\end{equation}%
where we have defined the polarization of the electrodes as $%
P_{l(r)}=(N_{\uparrow ,l(r)}-N_{\downarrow ,l(r)})/(N_{\uparrow
,l(r)}+N_{\downarrow ,l(r)})$. We have also introduced the quantity $P_{b}=2%
\mathcal{TU}/((\mathcal{T}^{2}+\mathcal{U}^{2}))=(T_{\uparrow
}^{2}-T_{\downarrow }^{2})/(T_{\uparrow }^{2}+T_{\downarrow }^{2})$, {which
is a measure for the spin-filter efficiency.} The quantity $P_{b}$\ equals
zero for spin-independent transmission coefficient and equals one for a
100\% spin-filter effect. %\textbf{%
%which is is not zero if the transmission coefficient of the barrier is
%spin-dependent.}
Eq. (\ref{Y(V)Sp1}) shows that in the presence of a spin-filtering effect ($%
P_{b}\neq 0$), the conductance depends on the relative angle $\beta $
between the magnetizations of the F electrode and the spin-filter barrier.

In the case of a $F/I_{sf}/F$ junction we obtain a general expression for
the spectral conductance
\begin{equation}
Y_{\alpha \beta }^{(FF)}=1+P_lP_r\frac{[\mathcal{T}^{2}\cos (\alpha -\beta )+%
\mathcal{U}^{2}\cos (\alpha +\beta )]}{(\mathcal{T}^{2}+\mathcal{U}^{2})}+P_b%
{[}P_r\cos \beta +P_l\cos \alpha ] \; ,  \label{Y(V)Sp3}
\end{equation}
In order to make a connection with the effect of tunnel magnetoresistance we
define the relative conductance change as
\begin{equation}
TMR=\frac{G_{00}-G_{0\pi}}{G_{0\pi}}\;.
\end{equation}
Thus, for the $F/I_{sf}/F$ junction one obtains
\begin{equation}
TMR=2\frac{P_r(P_l+P_b)}{1-P_r(P_l+P_b)+P_bP_l}\;.  \label{TMR1}
\end{equation}
If we assume that there is no spin-filter, \textit{i.e.} $P_b=0$ then Eq. (%
\ref{TMR1}) leads to the well-know Julliere's formula \cite{Julliere}.

Now we consider ferromagnetic electrodes with collinear magnetizations. We
distinguish two magnetic configurations: the parallel one \textbf{P},
\textit{i.e} $\beta =\alpha $ and the antiparallel configuration \textbf{AP}%
, $\beta =\alpha +\pi $. %%%%%%%
\begin{figure}[tb]
\includegraphics[width=\columnwidth]{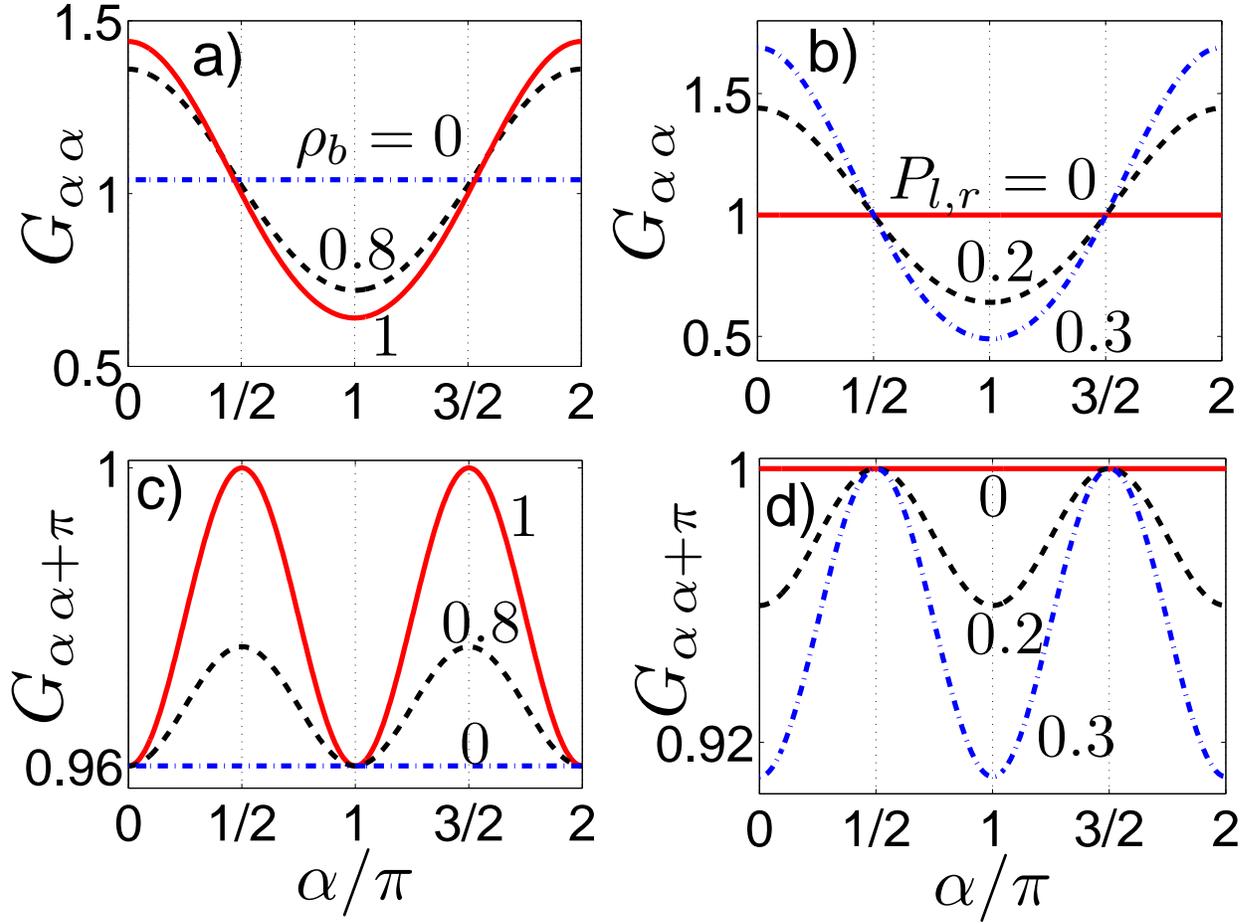} \vspace{-2mm}
\caption{The normalized conductance $G_{\protect\alpha ,\protect\beta }$ for
a $F_{l}/I_{sf}/F_{r}$ junction as a function of the magnetization
orientation. Panel (a): the parallel (\textbf{P}) configuration (\textit{i.e
} $\protect\beta =\protect\alpha $). Panel (b): the antiparallel (\textbf{AP}%
) configuration ($\protect\beta =\protect\alpha +\protect\pi $). The
polarizations in the left (right) electrodes $P_{l,r}$ are:$P_{l}=P_{r}=0.2$
(panels (a) and (c)). The parameter characterizing the spin-dependent
barrier transmission $P_{b}$ equals $P_{b}=1$ (panels (b) and (d)).}
\label{fig_fif}
\end{figure}
%%%%%%%
In Fig.(\ref{fig_fif}) we show the conductance of a junction with $%
P_{l}=P_{r}$ in these two cases. In the \textbf{P }configuration the
conductance is $2\pi $-periodic and, depending on the angle $\alpha $, it is
larger or smaller than in the non magnetic case. The largest value of the
conductance is obtained when the magnetizations of the electrodes and the
barrier are parallel. In the \textbf{AP} configuration the conductance is $%
\pi $-periodic and it is always smaller than in the non magnetic case. It is
interesting to note that in the case of a fully polarizing barrier $P_{b}=1$
and perpendicular magnetization of the ferromagnets with respect to $I_{sf}$
($\alpha =\pi /2$), the conductance equals \textbf{1} of a non-magnetic
junction for arbitrary value of the spin-polarization of the electrodes and
for both configurations \textbf{P} and \textbf{AP} (see panels (b) and (d)
in Fig.(\ref{fig_fif})).

%%%%%%%%

If the junction consists of two half-metallic ferromagnets (HM) then $%
P_{l,r}=1$ and from Eq. (\ref{Y(V)Sp3}) one obtains
\begin{equation}
Y_{\alpha \beta }^{(HM)}=1+\frac{[\mathcal{T}^{2}\cos (\alpha -\beta )+%
\mathcal{U}^{2}\cos (\alpha +\beta )]}{(\mathcal{T}^{2}+\mathcal{U}^{2})}%
+P_{b}{[}\cos \beta +\cos \alpha ]\;,  \label{Y(V)Sp3b}
\end{equation}%
It is clear from this expression that the conductance of a non-magnetic
barrier ($\mathcal{U}=0$) {vanishes if the magnetizations of the left and
right electrodes are antiparallel }($\beta =\alpha +\pi $).

If $\alpha =\beta =0$ and $\alpha =\beta =\pi ,$ the spectral conductance is
given by
\begin{equation}
Y_{00,\pi \pi }^{(HM)}=2(1\pm P_{b})  \label{Y(V)Sp3c}
\end{equation}%
As expected, the conductance of the $HM/I_{sf}/HM$ junction {with a barrier
impenetrable for one spin direction} vanishes if the magnetization in the
left and right electrodes are antiparallel with respect to the magnetization
of the barrier.

\subsection{Junctions with one superconducting electrode}

We consider now a $F/I_{sf}/S$ junction and calculate the conductance of the
system {using }Eqs. (\ref{Conductance}-\ref{Y(V)Sp}). {In the
superconducting electrode $\nu _{0,3,r}(\epsilon )=[\nu _{r}(\epsilon
+h_{r})\pm \nu _{r}(\epsilon -h_{r})]/2$, where $\nu _{r}(\epsilon
)=\epsilon /\sqrt{\epsilon ^{2}-\Delta ^{2}}$. Here $h_{r}$ is an effective
exchange field induced in the superconductor by the proximity of a thin
F-layer (as in a $F/I_{sf}/FS$ junction) or by the proximity of the magnetic
barrier $I_{sf}$ itself \cite{Sauls88a}. In this case $\nu _{0l}$, $\nu
_{0r} $ and $\nu _{3l}$ are even functions of $\epsilon $ and $\nu _{3r}$ is
an odd functions of $\epsilon $ (}in the quasiclassical approximation{).
Only the first and last terms of Eq. (\ref{Y(V)Sp}) contribute to the
integral in Eq. (\ref{Conductance}). Thus, the spectral conductance $%
Y_{\alpha \beta }(\epsilon )$ can be written as follows
\begin{equation}
Y_{\alpha \beta }(\epsilon )^{(FS)}=\nu _{0r}\left( 1+P_{b}P_{l}\cos \alpha
\right) \;.  \label{Y(V)Sps}
\end{equation}%
This equation, as well as Eq. (\ref{Y(V)Sp}), resembles the Slonczewski
formula\cite{Slonczewski}. It generalizes the latter for the case of a
superconducting electrode and a spin-filter barrier. %%%%%%%%%%%%%
\begin{figure}[tb]
\includegraphics[width=\columnwidth]{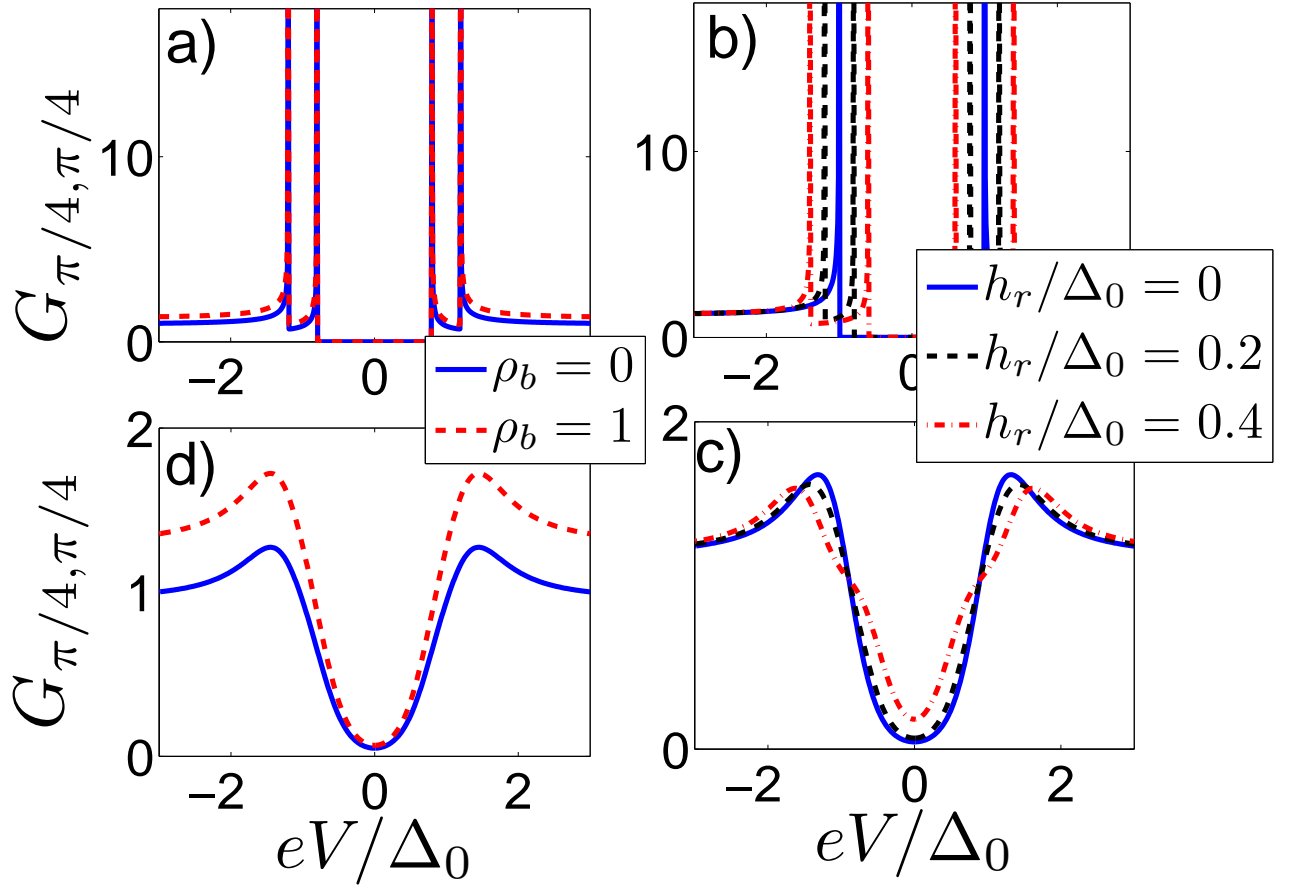} \vspace{-2mm}
\caption{The voltage dependence of the ratio $\tilde{G}=G_{\protect\pi /4,%
\protect\pi /4}/G_{0}$, where $G_{\protect\pi /4,\protect\pi /4}$ is the
normalized conductance for a $F/I_{sf}/S_{M}$ junction with fixed
polarization in the ferromagnetic electrode ($P_{l}=0.5$) and fixed angles ($%
\protect\alpha =\protect\beta =\protect\pi /4$), and $G_{0}$ is the
normalized conductance for a $F/I_{sf}/S_{M}$ junction with zero
polarization in the ferromagnetic electrode ($P_{l}=0$), zero exchange field
in the superconductor ($h_{r}=0$) and no spin-filter effect ($P_{b}=0$). In
panels (a) and (b) at zero temperature and in panels (c) and (d) at $%
T=0.2T_{c}^{0}$, where $T_{c}^{0}$ is the critical temperature for $h_{r}=0$%
. In panel (a) $h_{r}=0.2\Delta _{0} $ and $P_{b}=0,1$. In panel (b) $%
h_{r}=0,0.2,0.4\Delta _{0}$ and $P_{b}=0.8$. In panel (c) $h_{r}=0.2\Delta
_{0}$ and $P_{b}=0,0.8,1$. In panel (d) $h_{r}=0,0.2,0.4\Delta _{0}$ and $%
P_{b}=0.8$. By calculating the curves we have added a small $\protect\gamma %
=0.01\Delta _{0}$ damping factor.}
\label{fig_fis}
\end{figure}
%%%%%%%%%%%%%%%%
}

{In Fig.\ref{fig_fis} we show the conductance for the $F/I_{sf}/S$ junction
obtained from Eqs. (\ref{Conductance}, \ref{Y(V)Sps}). One can see the
splitting of the conductance peaks at $eV=\pm \Delta \pm h_{r}$ due to the
finite exchange field $h_{r}$ in the superconducting electrode. Note that by
increasing the temperature the peaks smeared out and are more difficult to
recognize. From Figs \ref{fig_fis}a and \ref{fig_fis}c one can see that the
values of $G$ in the normal state, \textit{i.e.} the asymptotic values for $%
V\gg \Delta $, depends on the polarization of the barrier, in accordance
with Eq. (\ref{Y(V)Sp1}). }

\section{Subgap Conductance in $N/I_{sf}/S$ Junctions: Effective Boundary
Conditions}

In the previous sections we have calculated the conductance of different
junctions in the tunneling limit. In other words, the Green's functions in
the left and right electrodes have not been corrected due to the proximity
effect. However, it is well known that the proximity effect in $N/S$
structures induces a condensate in the normal metal which causes a subgap
conductance in $N/S$ junctions \cite{Kastalsky91,Quirion02}. In order to
quantify the proximity effect we need boundary conditions that take into
account the spin-filtering at the $N/S$ barrier. Surprisingly, in spite of
many works on $N/S$ and $F/S$ structures such boundary conditions are absent
in the literature \cite%
{Zaitsev84,Shelankov84,Millis88,Tanaka97,Eschrig2000,Fogelstrom2000,Belzig09,Kalenkov09,Kawabata10}%
. In order to fill this gap we present here a heuristic derivation of the
boundary conditions at $S_{M}/I_{sf}/S_{M}$ interface which can also be used
for $N/I_{sf}/S$ or $S/I_{sf}/FS$ interfaces.

We consider the diffusive limit and write down the Usadel-like equation for
the Keldysh function $\check{g}^{K}$ in the $S_{M}$ electrodes
\begin{equation}
-D\partial _{x}(\check{g}\partial _{x}\check{g})^{K}-[i\hat{\tau}_{3}\otimes
(\hat{\sigma}_{0}\epsilon -\hat{\sigma}_{3}h(x))+\Delta \hat{\tau}%
_{2}\otimes \hat{\sigma}_{3},\check{g}^{K}]=0  \label{Usadel}
\end{equation}%
We assume that the exchange field $h$ differs from zero only in a thin
enough layer of thickness $d_{F}\ll \xi _{h}$. This allows us to integrate
Eq.(\ref{Usadel}) over the thickness $d_{F}$ considering the Green's
functions $\check{g}^{K}$ to be constant in this narrow layer. Performing
this procedure at $x>0$, we obtain for the ''spectral'' matrix current $%
\check{I}^{K}(\epsilon )$
\begin{equation}
\check{I}^{K}(\epsilon )\equiv (\check{g}\partial _{x}\check{g}%
)^{K}|_{x=d_{F}}=i\kappa _{h}[\hat{\tau}_{3}\otimes \hat{\sigma}_{3},\check{g%
}_{r(l)}]+\check{I}_{T}^{K}\text{ }  \label{BC1}
\end{equation}%
where $\kappa _{h}=hd_{F}/D$. The latter term describes the tunneling
current. The charge current density, for example, in the right
superconductor is given by
\begin{equation}
I=(\sigma _{r}/16e)\int d\epsilon \mathrm{Tr}\left\{ \hat{\tau}_{3}\otimes
\hat{\sigma}_{0}[\check{g}_{r}\partial _{x}\check{g}_{r}]^{K}\right\}
\label{TunCurrS}
\end{equation}%
This current equals the tunneling current given in (\ref{Current2}).
Therefore one can assume that
\begin{equation}
\check{I}_{T}^{K}=\text{ }\frac{\kappa _{T}}{\mathcal{T}^{2}+\mathcal{U}^{2}}%
\left[ \mathit{\check{\Gamma}}_{\alpha \beta }\,\check{g}_{l}\mathit{\check{%
\Gamma}}_{\alpha \beta }^{\dagger },\check{g}_{r}\right] ^{K}  \label{BC2}
\end{equation}%
where $\kappa _{T}=1/(\sigma _{r}R_{N\Box })$, $\sigma _{r}$ is the {%
conductivity} of the right $S$ electrode in the normal state, and $R_{N\Box }
$ is the interface resistance in the normal state per unit area.

Equations (\ref{BC1}) and (\ref{BC2}) represent the boundary conditions (BC)
for the Keldysh matrix function $\check{g}_{l}^{K}$. Equivalent equations
hold for the retarded (advanced) Green's functions, $\check{g}^{R(A)}$, if
the index $K$ is replaced by indices $R(A)$. We can then write a boundary
condition for the matrix Green's function $\check{g}${\ in a general form}
\begin{equation}
(\check{g}_{r(l)}\partial _{x}\check{g}_{r(l)})|_{x=0}=i\kappa _{h}[\hat{\tau%
}_{3}\otimes \hat{\sigma}_{3},\check{g}_{r(l)}]+\frac{\kappa _{T}}{\mathcal{T%
}^{2}+\mathcal{U}^{2}}\left[ \mathit{\check{\Gamma}}_{\alpha \beta }\,\check{%
g}_{l(r)}\,\mathit{\check{\Gamma}}_{\alpha \beta }^{\dagger },\check{g}%
_{r(l)}\right] \text{ }  \label{BC3}
\end{equation}%
%
%
%
%
%
%\begin{equation}
%(\check{g}_{\omega }\otimes \partial _{x}\check{g}_{\omega })|_{x=0}=i\kappa
%_{h}[\hat{\tau}_{3}\otimes \hat{\sigma}_{3},\check{g}_{\omega r(l)}]+\frac{%
%\kappa _{T}}{\mathcal{T}^{2}+\mathcal{U}^{2}}\left[ \check{\Gamma}_{\alpha
%\beta }\otimes \check{g}_{\omega l}\otimes \check{\Gamma}_{\alpha \beta
%}^{\dagger },\check{g}_{\omega r}\right] \text{ }  \label{BC3}
%\end{equation}
%
This condition generalizes the Kuprianov-Lukichev (K-L) BCs \cite{K-L} for
the case of spin-dependent transmission coefficients and in the presence of
an effective exchange field $h$ \footnote{%
Note, that Kupriyanov-Lukichev BCs, although quite useful, are
phenomenological ones. The authors of Ref.\cite{K-L} started with the
Zaitsev's microscopic BCs\cite{Zaitsev84}, but finally their derivation of
the BCs in the dirty limit was not rigorous. For a discussion of this
subject see Ref. \cite{Volkov97}}. Eq. (\ref{BC3}) is valid for the case
when the tunneling matrix elements $\mathcal{T}_{\uparrow }$ and $\mathcal{T}%
_{\downarrow }$ do not depend on momenta. In other words, no component of
the momentum is conserved by tunneling (diffusive interface). The physical
meaning of the BCs (\ref{BC3}) is rather simple. The first term stems from
finite exchange field in the vicinity of the $I_{sf}/S$ interface, while the
second term on the r.h.s. is due to the tunneling through the barrier with
spin-dependent transmission coefficients $\mathcal{T}_{\uparrow (\downarrow
)}.$ Note that in equilibrium, Eq. (\ref{BC3}) is also valid for the
Matsubara Green's functions $\check{g}_{\omega }$.

We emphasize that the above derivation of the BC Eq. (\ref{BC3}) cannot be
regarded as a microscopic derivation. However, these BCs give correct
physical results, and hence they can be used, for example, for the
calculations of the tunnel current in $S_{M}/I_{sf}/S_{M}$ junctions and for
the study the proximity effect in $I_{sf}/S_{M}$ and other systems.

% Note that the tunnel Hamiltonian method is known to give the
%correct results for the tunnel current which coincide with those obtained on
%the basis of microscopic Gor'kov's equations (see the books \cite%
%{KulikBook,BaroneBook} and references therein).

One can compare the BCs (\ref{BC3}) with those obtained earlier for
diffusive systems\cite{K-L,Belzig09}. In the nonmagnetic case, i. e. when
the matrix $\mathit{\check{\Gamma}}_{\alpha \beta }$ is a scalar $\mathcal{%
\Gamma }$ and $h=0$, Eq.(\ref{BC3}) coincides with the K-L BC \cite{K-L}.
% and with the first term at the
%right in Eq.(61) of Ref.\cite{Belzig09} which was obtained in the lowest
%approximation in the tunneling matrix element $\mathcal{T}^{2}$.
If $h\neq 0$, the first term at the r.h.s of Eq. (\ref{BC3}) coincides with
the third term in the r.h.s. of Eq.(61) of Ref.\cite{Belzig09}. Moreover, If
we assume that the magnetization vectors in the superconductors are parallel
to the $z$-axis then $\mathit{\check{\Gamma}}_{00}=\mathcal{T}+\mathcal{U}%
\hat{\tau}_{3}\otimes \hat{\sigma}_{3}$ and we obtain:
\begin{equation}
\left[ \mathit{\check{\Gamma}}_{00}\check{g}_{l}\mathit{\check{\Gamma}}%
_{00}^{\dagger },\check{g}_{r}\right] =\mathcal{T}^{2}\left[ \check{g}_{l},%
\check{g}_{r}\right] +\mathcal{U}^{2}\left[ \hat{\tau}_{3}\otimes \hat{\sigma%
}_{3}\check{g}_{l}\hat{\tau}_{3}\otimes \hat{\sigma}_{3},\check{g}_{r}\right]
+\mathcal{TU}[\{\hat{\tau}_{3}\otimes \hat{\sigma}_{3},\check{g}_{l}\},%
\check{g}_{r}]\;.  \label{comm}
\end{equation}%
We see that the last term proportional to $\mathcal{TU}$ corresponds to the
second term in the r.h.s of Eq.(61) of Ref.\cite{Belzig09}. However, as it
was shown in our previous work\cite{Bergeret12} this term does not
contribute to the Josephson current. The first correction to the current due
to the spin-filter is of the order $\mathcal{U}^{2}$ and described by the
second term in Eq. (\ref{comm}). The latter was neglected in Ref. \cite%
{Belzig09}. This term is essential if one needs to describe spin-filtering
effect. Just due to this term the Josephson current is zero if either $%
\mathcal{T}_{\uparrow }$ or $\mathcal{T}_{\downarrow }$ is zero \cite%
{Bergeret12}. Notice that the BC condition derived in Ref.\cite{Belzig09}
contains other terms which are product of three Green's functions, i.e. the
higher order terms in the expansion with respect to the tunneling
coefficients $\mathcal{T}$ and $\mathcal{U}$. The BCs (\ref{BC1},\ref{BC2})
also describe an interface between different materials with, for example,
different effective masses. In two recent works\cite{Burmistr12},\cite%
{Golubov12} the BCs at an interface between different materials in a
ballistic case were derived using another approach.

As an \textbf{example, we} use the derived boundary conditions (\ref{BC1},%
\ref{BC2}) to study the proximity effect in a simple $N/I_{sf}/S$ system
with a spin-filtering barrier. We assume a weak proximity effect and hence a
small amplitude of the condensate function $\check{f}_{N}$ induced in the
normal metal. We then can write $\check{g}_{N}=\hat{\tau}_{3}\otimes \hat{%
\sigma}_{0}+\check{f}_{N}$. The linearized BC (\ref{BC1}) acquires the form
\begin{equation}
\partial _{x}\check{f}_{N}|_{x=0-}=-\frac{\kappa _{T}}{\mathcal{T}^{2}+%
\mathcal{U}^{2}}\mathit{\check{\Gamma}}_{00}\check{f}_{S}\mathit{\check{%
\Gamma}}_{00}^{\dagger }=-r\kappa _{T}f_{S}.\text{ }  \label{PE1}
\end{equation}%
where $\check{f}_{S}^{R(A)}=\hat{\tau}_{2}\otimes \hat{\sigma}%
_{3}f_{S}^{R(A)},$ $f_{S}^{R(A)}=\Delta /\sqrt{\Delta ^{2}-(\epsilon \pm
i\gamma )^{2}}$ is the amplitude of the quasiclassical anomalous (Gor'kov's)
Green's function in the $S$ superconductor and $\check{\Gamma}_{00}=\mathcal{%
T}+\mathcal{U}\hat{\tau}_{3}\otimes \hat{\sigma}_{3}$. We have defined the
spin-filter parameter as $r=(2\mathcal{T}_{\uparrow }\mathcal{T}_{\downarrow
})/(\mathcal{T}_{\uparrow }^{2}+\mathcal{T}_{\downarrow }^{2})$. The latter
is related to the spin-filter efficiency $P_{b}$ of the spin-filter barrier
by the expression $r=\sqrt{1-P_{b}^{2}}$. For a barrier transparent only for
one spin direction $r=0$, while for a non-magnetic one $r=1$.

The condensate in the normal metal has the same matrix structure as in S, $%
\check{f}_{N}=\hat{\tau}_{2}\otimes \hat{\sigma}_{3}f_{N}$, where the
amplitude $f_{N}$ is found from the linearized Usadel equation
\begin{equation}
\partial _{xx}^{2}f_{N}-\kappa _{\epsilon }^{2}f_{N}|^{R(A)}=0  \label{PE2}
\end{equation}%
complemented with the BC (\ref{PE1}). The solution of Eq.(\ref{PE2}) can be
easily written
\begin{equation}
f_{N}(x)=r(\kappa _{T}/\kappa _{\epsilon })f_{S}\exp (\kappa _{\epsilon
}x)\;,  \label{PE3}
\end{equation}%
where $\kappa _{\epsilon }^{2}|^{R(A)}=\mp 2i\epsilon /D$. Thus, the
amplitude $f_{N}$ of the induced condensate is proportional to spin-filter
parameter $r$. In particular the proximity effect is completely suppressed
if the barrier is fully spin-polarizing ($r=0$). Although this result is
quite obvious, it has not been obtained in any previous work.

Now consider the case when a voltage difference $V$ is applied to the $%
N/I_{sf}/S$ junction. Such a situation was studied both experimentally \cite%
{Kastalsky91,Quirion02} \footnote{%
To be more exact in the experiments of Refs.\cite{Kastalsky91,Quirion02} a
highly doped semiconductor-Shottky barrier-superconductor structure was
explored} and theoretically \cite%
{Wees92,Beenakker92,Volkov92,VZK93,Nazarov94,Bergeret12a}. It was observed
that a zero-bias peak arises in the voltage dependence of the conductance.
The origin of these peak is the induced condensate in the normal metal. In
this case, the tunnel current consists not only of the usual quasiparticle
current, but also of the current which is proportional to product of the
condensate amplitudes in $S$ and $N$ electrodes and to the applied voltage $%
V $. We now calculate this conductance in the presence of a spin-filter
barrier.

The subgap current, which describes charge transfer below the
superconducting gap, can be obtained from Eq.(\ref{Current1})
\begin{equation}
I_{sg}=(32eR_{N})^{-1}\int d\epsilon \mathrm{Tr}\left\{ n_{V}\hat{\tau}%
_{0}\otimes \hat{\sigma}_{0}[(\check{f}_{l}^{R}+\check{f}_{l}^{A})(\check{f}%
_{r}^{R}+\check{f}_{r}^{A})]\right\} \; .  \label{sg}
\end{equation}%
{At low energies} ($\epsilon \sim eV\ll \Delta $) one has $\check{f}%
_{r}^{R}\approx \check{f}_{r}^{A}\approx -\hat{\tau}_{2}\otimes \hat{\sigma}%
_{3}$ and $\check{f}_{l}^{R(A)}\approx -r[(\kappa _{T}\xi _{T})(1\pm i)\sqrt{%
2T/|\epsilon |}]\hat{\tau}_{2}\otimes \hat{\sigma}_{3}$, $\xi_T=\sqrt{D/(2T)}
$. For the normalize differential conductance, $\mathit{G}%
_{sg}=R_{N}dI_{sg}/dV$, we obtain
\begin{equation}
\mathit{G}_{sg}=\sqrt{1-P_b^2}(\kappa _{T}\xi _{T}/16)J(V) \; ,  \label{Gsg}
\end{equation}%
where we have use the fact that $r=\sqrt{1-P_b^2}$. Here $J(V)=\int
d\epsilon \lbrack dn_{V}/d(eV)]\sqrt{2T/\epsilon }$. For $V=0$ one has $%
J(0)\approx 3.41$, whereas in the limiting case of low temperatures ($V\gg T$%
), one obtains $J(V)\approx \sqrt{2T/eV }$. It is clear from Eq. (\ref{sg})
that {spin-filtering suppresses} the subgap conductance. %%%%%%
\begin{figure}[tb]
\includegraphics[width=\columnwidth]{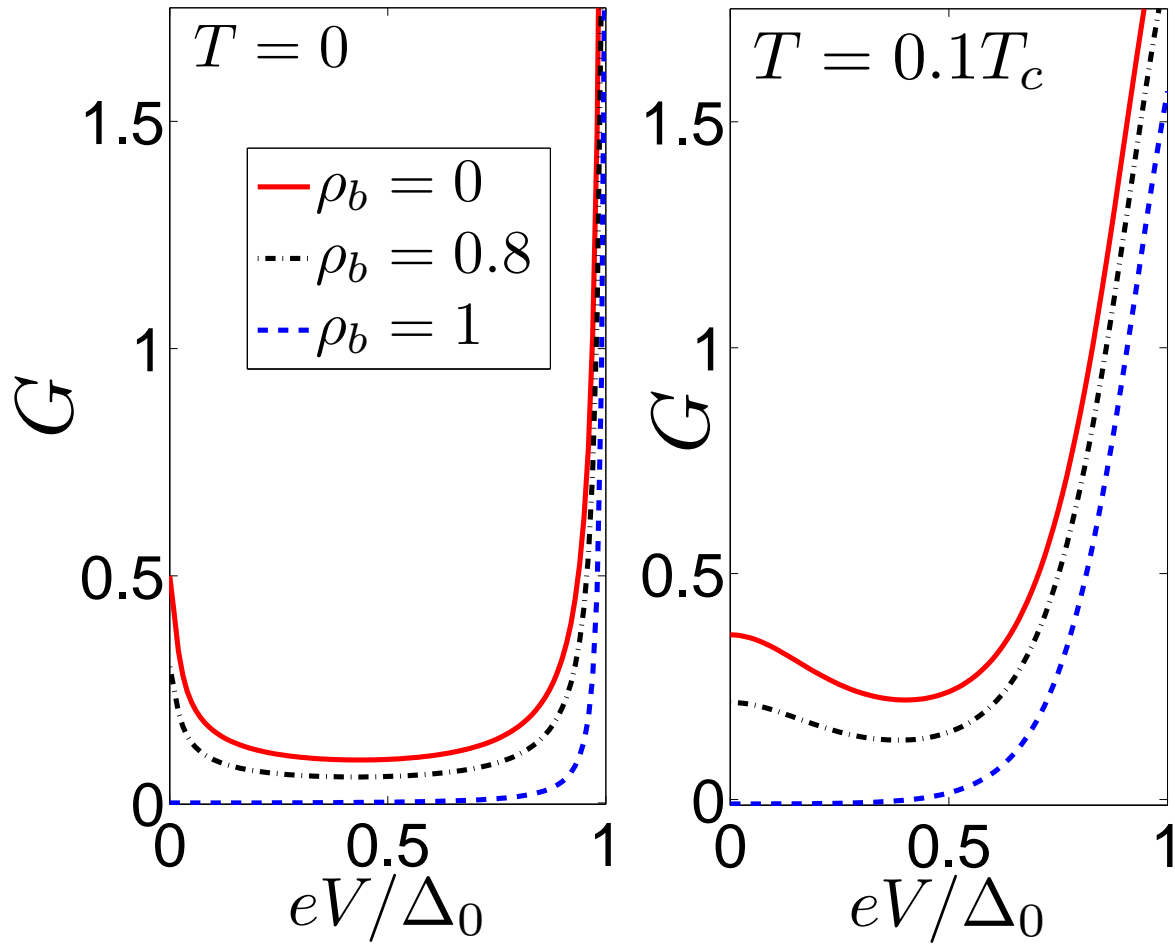} \vspace{-2mm}
\caption{The voltage dependence of the normalized conductance $G$, for a N/I$%
_{sf}$/S$_{M}$ junction, calculated from Eqs.(\ref{Y(V)Sp},\ref{sg}). We set
$\protect\kappa_T\protect\sqrt{D/\Delta_0}=0.25$ and $T=0 $ in the left
panel and $T/T_c=0.1$ in the right one. By calculating the curves we have
added a small $\protect\gamma= 0.01\Delta_0$ as damping factor.}
\label{fig4}
\end{figure}
%%%%%
In Fig.(\ref{fig4}) we plot the voltage dependence of the normalized
differential conductance taking into account the contribution of the
quasiparticle and subgap currents given in Eq.(\ref{Y(V)Sp}) and Eq.(\ref{sg}%
) respectively.

\section{Conclusions}

We have studied the transport properties of a generic $X_{l}/I_{sf}/X_{r}$
junction with a spin-filter tunneling barrier $I_{sf}$. The electrodes $%
X_{l,r}$ can be a normal metal, a ferromagnet, or a superconductor with or
without a build-in exchange field. We have derived a general expression for
the tunneling conductance, Eq. (\ref{Conductance}), which is valid for
arbitrary value of the exchange fields and the angles between the
magnetizations. This expression generalizes the well known results for
normal multilayer systems with collinear magnetization and shows how the
conductance depends on the mutual orientation of the magnetization of the
electrodes and the magnetic barrier, the spin-filter parameter and the
spin-dependent density of the states in the normal and superconducting
electrodes. {We also} have derived new boundary conditions for the
quasiclassical\ Green's functions taking into account the spin-filter
effect. By using these boundary conditions we have studied the proximity
effect in $N/I_{sf}/S$ system and showed that spin-filtering suppresses the
amplitude of the condensate in the normal layer. In particular we show that
the sub gap conductance of the $N/I_{sf}/S$ junction is suppressed due to
the spin-filter effect by a factor $\sqrt{1-P_b^2}$, where $P_b$ is the
spin-filter efficiency parameter.

\section*{Acknowledgements.}

The work of F.S.B and A. V. was supported by the Spanish Ministry of Economy
and Competitiveness under Project FIS2011-28851-C02-02 and the Basque
Government under UPV/EHU Project IT-366- 07. A. F. V. is grateful to the
DIPC for hospitality and financial support. F.S.B thanks Prof. Martin
Holthaus and his group for their kind hospitality at the Physics Institute
of the Oldenburg University.

\end{document}